\documentstyle [12pt]{article}
\font\csbig=rptmr at 14pt

\begin{document}

\hfill {WM-93-114}

\hfill December 1993

\vskip 1in   \baselineskip 24pt
{
\csbig

   \bigskip
   \centerline{Photoproduction of very light gluinos}
}
 \vskip .8in

\centerline{C.E. Carlson and Marc Sher}  \bigskip
\centerline {\it Physics Department, College of William and Mary}
\centerline {\it Williamsburg, VA 23187, USA}
\vskip 1in

{\narrower\narrower
Current experiments allow the possibility of gluino
masses below about 600 MeV if the lifetime of the gluino is
longer than 100 picoseconds.  If the mass and lifetime are in
this window, then photoproduction of pairs of gluino-gluon
bound states can provide a means to observe them.  The cross
section is large enough that the window can be fully explored,
up to lifetimes exceeding a microsecond,  at high luminosity
electron accelerators.

}

\newpage

\section{Introduction}

\def\tg{\tilde{g}}

In recent years, an apparent inconsistency between the
value of
$\alpha_s$ at low energies and that at the mass of the $Z$ has
led to a revival of interest in the possibility of very light
gluinos \cite{clavelli}.  Although the latest data seems to be
consistent, within errors, both with and without light
gluinos, the possibility that the gluino is extremely light
needs to be thoroughly explored.

 In 1987,  the UA1
Collaboration \cite{uaone} published a detailed analysis of the
experimental searches for gluinos.  They found
three allowed windows in the gluino-squark mass plane:
(i) gluino masses below approximately 600 MeV and squark mass
above something like 100 GeV, (ii) a triangular shaped window for
gluino masses between 2.5 and 4.0 GeV and squark masses between
100 and 400 GeV and (iii) a window for gluino masses between 2.0
and 5.0 GeV and squark masses in the TeV range.  These windows
are all controversial;
 looking at the long listings in the Particle
Data Group table \cite{pdg} for gluino masses will show the
extent of the controversy. In this letter, we will focus on
the most intriguing window---gluino masses below 1 GeV.

Because of R-parity, gluinos will
always be produced in pairs.  Once produced, they will either
combine with each other into a $\tg\tg$ state, a
``gluinoball,'' and then annihilate quickly into hadrons, or
else they will hadronize with gluons or quarks into a
``glueballino'' ($g\tg$) or a ``gluino hybrid'' or ``hybridino''
($\tg q\overline{q}$) state.  In the case of the single gluino
hadron, the lightest resulting state will be long-lived, since
the gluino will decay into a
$q\overline{q}\tilde{\gamma}$ via squark exchange with a
lifetime approximately given by \cite{hab}

\begin{equation}
\tau \sim 3\times 10^{-12}\ {\rm sec.} \left( 1\ {\rm GeV} \over
\widetilde{M} \right)^5  \left( m_{\rm squark}\over
m_W \right)^4,
\end{equation}
 where $\widetilde{M}$ is
the mass of the $g\tg$ or $\tg q\overline{q}$ state.  For
squark masses between
$50$ and $2000$ GeV, this gives lifetimes ranging from a
picosecond to a microsecond.

Present limits on light gluino masses come from searches
for $\tg\tg$ gluinoballs in radiative heavy
vector meson decays (e.g., $\Upsilon \rightarrow \gamma +$
gluinoballs). Such processes have the advantage of being
completely independent of the gluino lifetime, squark masses,
etc. The best bound comes from CUSB
\cite{cusb}, who exclude gluinos with masses between $600$ and
$2200$ MeV coming from radiative $\Upsilon$ decays.  The lower
bound comes from the low detection efficiency of low
multiplicity final states, and is quite
uncertain.\footnote[1]{The bound refers to half the gluinoball
mass.  The gluinos could conceivably be somewhat heavier or
lighter, or even massless.}  The bound has been
criticized
\cite{clav} since the determination of the expected branching
ratio
\cite{wu} is very strongly dependent on the value of the wave
function at the origin of the gluino-ball, and is thus quite
model-dependent, and because decays into more expectable things
such as $\gamma +
\eta^\prime$ and $\gamma +$ glueball have also not been seen.

Lifetime limits come from searches for
$g\tg$ glueballinos or a $\tg q\overline{q}$ states in beam dump
experiments.  Such experiments have conclusively ruled out
\cite{dump} gluinos with lifetimes less than $10^{-10}-10^{-11}$
seconds.  If a $\tg q
\overline{q}$ {\it charged} state has a lifetime greater
than
 $\sim 10^{-10}$ seconds, then it would have been detected
\cite{terry} in hyperon beam experiments. However, if the mass
of the $\tg q\overline{q}$ state is sufficiently greater than the
glueballino then it will decay
strongly into the glueballino, and such a bound would not
be relevant.\footnote[2]{In several models, the $\tg
d\overline{u}$ state will be sufficiently heavy to decay
into a glueballino and one or two pions; however, in most
of these models, the $\tg s\overline{u}$ state will not be
able to decay strongly into a glueballino and a kaon.  In
these models, the uncertainties in the masses are
sufficiently large that such a decay cannot be excluded;
furthermore, the $W$-mediated decay of the $\tg
s\overline{u}$ into a glueballino and a charged pion will
occur with a lifetime of approximately $10^{-10}-10^{-11}$
seconds, and thus might not be detected in the hyperon
beam.}
 We conclude that there may still be a window for
gluino masses less than approximately 1000 MeV and
lifetimes between $100$ picoseconds and a microsecond if
the lightest gluino containing hadron is a $g\tg$ glueballino.
In this letter, we will propose an experiment that could close
this window---or find the gluino.

In order to detect the decays of a neutral particle whose
lifetime could be as long as a microsecond, one needs to
produce them with very little kinetic energy (i.e. a
relatively low energy machine) and with a very high
luminosity.  We will consider the photoproduction of light
gluinos off a proton target at a high luminosity electron
accelerator.

\section{Photoproduction of light gluinos}

The
relevant diagrams are shown in Fig. \ref{fig1}.  We will first
consider the production rate of light gluinos, and then
discuss signatures in the next section.

The square of the matrix element of the diagrams of Fig.
\ref{fig1} is given by

\begin{equation}
\big| M    %{\cal M}
\big|^2={64g^4_s e^2  \over -\hat{u}\hat{s} r^4}
\big[(r^2-2\Delta^2)(\hat{s}^2+\hat{u}^2+2r^2\hat{t})+8r^2((p\cdot
\Delta)^2+(p'\cdot\Delta)^2)\big]
\end{equation}
where $r^2$ is the invariant mass of the gluino pair,
$\Delta$ is half the difference between the four-momenta of
the gluinos, and $p$ and $p'$ are the four-momenta of the
initial and final quarks, respectively.  We have omitted a
factor $e^2_q$ for the quark charge which we shall restore
before our final calculation.  In integrating over phase space,
it is convenient to first write the integrals in covariant form,
pick the $\vec r =0$ frame, do the integrations over gluino
momenta,  and re-express the result in covariant form before
doing the integral over the outgoing quark directions in the
subprocess center of mass.  The resulting cross-section is given
by

\begin{equation}
{d{\hat \sigma}\over
d\epsilon}={64 \alpha \alpha_s^2 \over 3}
{\left( 1-\epsilon-\tilde \mu^2 \right)
\sqrt{1-{4\tilde \mu^2\over 1-\epsilon+ \mu_q^2}}\over
(1-\epsilon)^2}  \left(2\left( (1-\epsilon)^2+\epsilon^2
\right)\log{1+\beta\over 1-\beta}+4\epsilon-3\epsilon^2\right)
\end{equation}
where $\tilde\mu$ and
$\mu_q$ are the gluino mass and target
quark mass scaled by
$\sqrt{\hat{s}}$ and and $\epsilon$ is twice the outgoing
quark energy (in the subprocess center of mass) scaled by
$\sqrt {\hat s}$.  Here,
$\beta=\sqrt{1-4\mu_q^2/\epsilon^2}$ is
the final quark velocity.  We kept the mass of the
final state quark only when necessary to
avoid infrared singularities---letting the quark mass vary from
$300$ to $1000$ (!) MeV will give an indication of the
sensitivity of the calculation to this mass.   The limits of
$\epsilon$ integration are from
$2 \mu_q$ to $1-(4\tilde \mu^2-\mu_q^2)$.

After we obtain the subprocess cross section,  we must
embed the target quark in a proton and integrate
over the allowed range of $\hat s$. For various incoming photon
energies and various particle masses,  we obtain the
cross-sections shown in Figure \ref{fig2}.  Some details follow.

We fold the subprocess cross section with the distribution
functions of the quark in a proton,

\begin{equation}
\sigma = \int dx\, \sum_q e_q^2 f_q(x) \hat \sigma(\hat s)
= \int dx\, \hat \sigma(\hat s) F_{2p}(x)/x.
\end{equation}
where $F_{2p}$ is the proton electromagnetic
structure function and the scale (i.e., $Q^2$,  where
$Q$ is some relevant momentum transfer) dependence of $f_q$ is
tacit.  We used the up-to-date CTEQ distributions \cite{CTEQ},
specifically CTEQ1L,  for Fig. \ref{fig2}.
\footnote[3]{In addition,  some
old but simple distribution functions \cite{CH83} were used for
calibration purposes.  The results using the Ref. \cite{CH83}
distributions were about 30\% below the CTEQ results over
most of the plotted range,  although they were slightly
higher  very near threshold.  This mirrors the behavior of the
distributions functions in $x$,  since the closer we are to
threshold in our process the higher the average $x$ must be,
and the Ref. \cite{CH83} distributions are higher (and
actually fit the limited amount of non-resonance region data
better) at high
$x$,  whereas the CTEQ distributions are higher (and fit the data
better) at $x < 0.75$.  The average $x$ for
the top curve in Fig. \ref{fig2} is unity at
threshold,  passes 0.75 at $\omega = 6$ GeV,  and is 0.39 at the
right hand edge.}

The relation between $x$ and $\hat s$ at high energy,  where
one can neglect masses,  is clear.  One has $x = \hat s /s$.
We have used a modification of this just to ensure that the
threshold points of $\hat s$ and $s$ are maintained,
namely

\begin{equation}
x = \left( \sqrt{\hat s} - m_q \over \sqrt{s}- m_N \right)^2
\end{equation}
where $m_N$ is the nucleon mass.  This has little effect except
near threshold where the cross section is small anyway.

We envision each gluino within a glueballino (a
bound state of gluinos with gluons) so that the mass
necessarily produced is at least that of the glueballino,  and in
evaluating our formulas we have interpreted $\tilde m$ as the
glueballino mass.  Regular glueballs are not massless even
though the gluon is,  and we anticipate that the glueballino will
be in the same mass range,  namely a mass of about 1.5 GeV for
the lightest example (see, e.g., the lattice gauge
results reported in \cite{MT89}).  Our cross section is
sensitive to this mass,  as may be seen from the figures,  where
we present results for $\tilde m$ being both 1.0 and 1.5 GeV.
The cross section is in contrast insensitive to changes in the
quark mass.

\section{Signatures of gluino production}

A signature of a gluino in the mass and lifetime range we are
considering is that it appears in certain aspects as a long lived
particle and in other aspects as a short lived particle.  The
particle is in fact long lived so that there should be a
noticeable gap between its production point and decay point.
For lifetimes near the low end of the $10^{-10}$ to
$10^{-6}$ second range and
a roughly 10 GeV incoming photon beam,  many of the gluinos
produced will have a measurable gap before decay, while for
lifetimes near the high end of the range,  some gluinos (at
least 1\%) will decay in the detector.

The gluino will decay into a photino plus non-supersymmetric
particles and the photino will exit undetected and with its
energy undetermined.  The ordinary matter from the glueballino
decay will therefore have a variable energy and will appear
like a strongly unstable particle  with a wide width.  The
apparent width of the decay will of course not have a lorentzian
shape,  but this may not be apparent if the statistics are
limited in a first experiment.

The cross section scale is of the order of nanobarns.  For a
photon luminosity of $10^{34}$ cm$^{-2}$sec$^{-1}$,  a number
pertinent to the large acceptance spectrometer at CEBAF,  a
nanobarn gives ten events per second.  Higher energy machines will
be less suitable for detecting gluinos with long lifetimes due to
both the time dilation factor as well as lower luminosity.

A final state such as four charged pions,
would, to judge from the decays of other particles in this mass
range \cite{pdg},  have a sufficient branching ratio to
give several gluino counts per hour if they are there.  This
final state would be easily detectable and there seems to be
no other particle that could produce it
with a significant apparent width,  and yet have its decay point
significantly separated from its production location.
\footnote[4]{
One could also look for two and three body final states,  which
may also appear with significant apparent width noticeably far
from the interaction region,  although the backgrounds would be
larger.}

\bigskip

{\underline{Acknowledgments}}.
We wish to thank L. Clavelli,  T. Goldman,  H. Haber,  and L.
Weinstein for useful comments.  We also thank the NSF for
support under Grant NSF-PHY-9306141.

\newpage

\begin{figure}

\vglue 0in  %test ratio using CH
\hskip 0.5in {\special{picture one scaled 1000}} \hfil

\caption{Feynman diagrams for photoproduction of gluino pairs
via photoproduction off a quark.  The corkscrew line is a gluon
and the lines labeled $k_1$ and $k_2$ are the gluinos.}
   \label{fig1}
\end{figure}

\begin{figure}
\vskip 1.4in

\vglue 0in  %sigma for several mq and mtw
\hskip 0.5in {\special{picture sigma scaled 1000}} \hfil

\caption{ The total cross section for photoproduction of
    glueballino pairs.  The upper curves are both for glueballino
    mass (inserted for $\tilde m$ in our formulas) of 1.0 GeV and
    the lower curves have glueballino mass 1.5 GeV.  The solid
    curves are for quark mass $m_q$ of 0.3 GeV and the dashed
    curves use $m_q$ = 1.0 GeV.  The CTEQ1L quark
    distributions at their benchmark of $Q^2 = 4$ GeV$^2$ were
    used for this figure.  As seen,  the results are
    sensitive to gluino mass but not to quark mass.}
  \label{fig2}
\end{figure}


\newpage

\begin{thebibliography}{99}

\bibitem{clavelli} L. Clavelli, Phys. Rev. D {\bf 46},
2112 (1992).

\bibitem{uaone} UA1 Collaboration, C. Albajar {\it et al.}, Phys.
Lett. B {\bf 198}, 261 (1987).

\bibitem{pdg} Particle Data Group, K. Hikasa {\it et al.}, Phys.
Rev. D {\bf 45}, S1 (1992).

\bibitem{hab} H. E. Haber and G. L. Kane, Phys. Reports {\bf 117},
75 (1985).

\bibitem{cusb} CUSB Collaboration, Phys. Lett. B {\bf 186},
233 (1987).

\bibitem{clav} L. Clavelli, P.W. Coulter and K. Yuan, Phys. Rev.
D {\bf 47}, 1973 (1993).

\bibitem{wu} W.-Y. Keung and A. Khare, Phys. Rev. D {\bf 29},
2657 (1984);

T. Goldman and H. Haber, Physica D {\bf 15}, 181 (1985).

\bibitem{dump} R.C. Ball {\it et al.}, Phys. Rev. Lett. {\bf 53},
1314 (1984); F. Bergsma {\it et al.}, Phys. Lett. {\bf 121B},
429  (1983); G.R. Farrar, Phys. Rev. Lett. {\bf 55}, 895 (1985); M.
Cooper-Sarkar {\it et al.}, Phys. Lett. {\bf 160B}, 212 (1985).

\bibitem{terry} T. Goldman, Phys. Lett {\bf 78B}, 110 (1978);

G.R. Farrar and P. Fayet, Phys. Lett. {\bf 76B}, 442, 575 (1978).

\bibitem{CTEQ} J. Botts {\it et al.},  Phys. Lett. B {\bf 304},
159 (1993).

\bibitem{CH83} C.E. Carlson and T.J. Havens,  Phys. Rev. Lett. {\bf
51}, 261 (1983).

\bibitem{MT89} C. Michael and M. Teper,  Nucl. Phys. B {\bf 314},
347 (1989).

\end{thebibliography}
\end{document}